\documentclass[%
 reprint,
superscriptaddress,
 amsmath,amssymb,
 aps,
]{revtex4-1}
\usepackage{ragged2e} 
\usepackage{booktabs}
\usepackage{multirow}
\usepackage{rotating}
\usepackage{subcaption}

\usepackage{float}
\usepackage{caption}
\usepackage{bm} 
\usepackage{graphicx}
\usepackage{dcolumn}
\usepackage{bm}
\usepackage{hyperref}

\begin{document}

\preprint{APS/123-QED}

\title{Dynamics of thin accretion disks and accretion around a charged-PFDM black hole}
\author{Taiyang Zhang}
\affiliation{%
 College of Physics,Guizhou University,Guiyang,550025,China
}%

\author{Zhongyuan Qin}
\affiliation{Department of Physics, Guizhou University, Guiyang, 550025, China}

\author{Qian Feng}
\affiliation{Department of Physics, Guizhou University, Guiyang, 550025, China}

\author{Zheng-Wen Long}%
\email{zwlong@gzu.edu.cn (corresponding author)}
\affiliation{%
 College of Physics,Guizhou University,Guiyang,550025,China
}%


\begin{abstract}
This paper investigates the dynamical behavior of steady spherical accretion onto a static, magnetically charged black hole embedded in a perfect fluid dark matter (PFDM) background. Using the shadow observations of M87* from the Event Horizon Telescope (EHT), we establish constraints on the parameter space for the magnetic charge and the PFDM parameter. Within this constrained range, we analyze the orbital dynamics of particles in a thin accretion disk surrounding the black hole and find that the black hole parameters significantly influence the effective potential, angular velocity, specific energy, and specific angular momentum of the particles. Subsequently, we calculate the radiative energy flux, temperature profile, and observed spectrum of the disk. Our results show that, while the local radiative flux and temperature at a given radius are lower for the charged-PFDM black hole compared to a Schwarzschild black hole, its overall radiative efficiency and total luminosity are higher. Finally, we explore the spherically symmetric, steady-state accretion process around the black hole, revealing how the parameters govern how the fluid velocity, density profile, and black hole mass accretion rate are influenced.
\end{abstract}

\maketitle


\section{Introduction}

Black holes, as one of the most fascinating predictions of general relativity, have transitioned from theoretical constructs to observable astrophysical objects over the past decade. The groundbreaking detections of gravitational waves from black hole mergers by LIGO and the first image of the supermassive black hole M87* captured by the Event Horizon Telescope (EHT) have ushered in a new era of black hole astrophysics \cite{LIGOScientific:2016emj, EventHorizonTelescope:2019dse}. These observations not only confirm the existence of black holes but also provide unprecedented opportunities to test gravitational theories in the strong-field regime. However, general relativity predicts the existence of spacetime singularities within black holes \cite{EventHorizonTelescope:2019ths}, where classical physics breaks down, signaling the need for more fundamental theories.

Nonlinear electrodynamics (NED), originally proposed by Born and Infeld in 1934 to resolve the infinite self-energy problem of point charges in classical electrodynamics, has emerged as a promising framework for addressing black hole singularities \cite{Gibbons:2001gy,Born:1934ji, Born:1934gh}. Coupling nonlinear electrodynamics with general relativity makes it possible to construct black hole solutions in which the central singularity is regularized and replaced by a finite curvature core \cite{Gunasekaran:2012dq, Bronnikov:2000vy, Bronnikov:2017sgg, Rincon:2021hjj, Bronnikov:2022ofk}. This line of thought regained widespread attention in the context of string theory, when it was realized that the effective action of open strings on D branes can be precisely described by nonlinear electrodynamics \cite{Gibbons:2001gy, Fradkin:1985qd, Seiberg:1999vs}. Over the past decades, numerous regular black hole solutions have been constructed within the framework of general relativity coupled to NED, with applications ranging from cosmological models to astrophysical phenomena \cite{DeLorenci:2002mi, Novello:2003kh, Novello:2008ra, Novello:2008xp, Xiang:2013sza, Balart:2014cga, Culetu:2014lca}. Extending these studies to rotating black holes, Lombardo et al. \cite{CiriloLombardo:2004qw} investigated Kerr-Newman-type solutions using the Newman-Janis algorithm and related methods \cite{Newman:1965tw, Azreg-Ainou:2014aqa, Azreg-Ainou:2014nra, Azreg-Ainou:2014pra}. Subsequently, Ghosh \cite{Ghosh:2021clx} further considered spacetimes with reflection symmetry and scalar polynomial singularities at $r=0$ \cite{Hawking:1973uf}, demonstrating that for certain ranges of the NED charge parameter, the black hole possesses both inner and outer horizons.

Perfect fluid dark matter (PFDM) provides an effective phenomenological description of dark matter that successfully reproduces the asymptotically flat rotation curves observed in spiral galaxies  \cite{Kiselev:2003ah, Kiselev:2002dx, Guzman:2000zba, Rahaman:2010xs, Potapov:2016obe}. Although dark matter may not strictly satisfy the perfect fluid condition, this model has been widely adopted due to its simplicity and consistency with observational data \cite{Rizwan:2018rgs, Ndongmo:2021how, Xu:2018wow}. Recently, Vachher et al. \cite{Vachher:2024ldc} investigated strong gravitational lensing by a static spherically symmetric magnetically charged black hole coupled to PFDM, while quasinormal modes of such black holes have also been studied \cite{Tan:2025usr}. The EHT observations of M87* and Sgr A* have imposed stringent constraints on the parameter space of these black hole models \cite{EventHorizonTelescope:2022wkp, EventHorizonTelescope:2021dqv}, providing a solid foundation for further theoretical investigations. 

Accretion processes around black holes offer one of the most powerful probes for testing general relativity and its modifications. Among various accretion models, the thin accretion disk model developed by Shakura-Sunyaev and Novikov-Thorne \cite{Shakura:1972te, Thorne:1974ve, Page:1974he, Novikov:1973kta} has become the standard framework for studying black hole accretion systems due to its observational accessibility and theoretical tractability. The innermost stable circular orbit (ISCO) serves as the inner boundary of the thin disk and determines the radiative efficiency—the fraction of rest-mass energy converted into radiation. Previous studies have demonstrated that accretion disk properties, including energy flux, temperature distribution, and emission spectrum, can reveal fundamental characteristics of the central black hole and test alternative gravity theories \cite{Li:2004aq, Pun:2008ua, Harko:2010ua, Chakraborty:2014eha, Bambi:2015kza, He:2022lrc}. However, most existing studies focus either on the thin disk structure or on spherical accretion separately, leaving a comprehensive investigation of both accretion regimes within a unified dark matter-black hole framework relatively unexplored. This study aims to fill this gap by systematically investigating both thin disk accretion and spherical accretion onto a magnetically charged black hole embedded in a PFDM background. The motivation for including both accretion regimes is twofold: First, thin accretion disks are relevant for high-luminosity active galactic nuclei where matter possesses sufficient angular momentum, while spherical accretion describes more diffuse, low-angular-momentum environments that may better represent dark matter accretion \cite{Yuan:2014gma}. Second, comparing these two regimes allows us to predict the full spectrum of accretion signatures—from the high-energy radiation of thin disks to the dynamical evolution of black hole mass through dark matter accretion. By combining these complementary perspectives, we can provide a more complete characterization of how magnetic charge and PFDM parameters influence observable phenomena, thereby offering theoretical predictions that can be tested by future multi-messenger observations of supermassive black holes.

In this paper, we investigate several aspects of the charged-PFDM black hole proposed in \cite{Vachher:2024ldc}. In Section \ref{sec:2}, we impose constraints on the magnetic charge and PFDM parameters using EHT shadow data from M87*. In Section \ref{sec:3}, we derive the effective potential, equations of motion, and metric components. In Sections \ref{sec:4}, we discuss the radiation flux, temperature profile, radiation efficiency, observed luminosity, and emission efficiency of a thin accretion disk. In Sections \ref{sec:5}, we examine the key parameters characterizing steady-state spherical accretion and determine how they influence the fluid velocity, density profile, and mass growth rate of the black hole. The final section provides a summary of our findings. In this study, we adopt a unit system in which $8\pi G = c = 1$ and the black hole mass $M=1$. Under this convention, the Einstein field equations reduce to $G_{\mu\nu} = T_{\mu\nu}$.

\section{Black hole shadow constraints from M87*}
\label{sec:2}

The line element of charged-PFDM black hole metrics is provided by \cite{Vachher:2024ldc}:

\begin{equation}
ds^2 = -f(r)\,dt^2 + \frac{1}{f(r)}\,dr^2 + r^2\left(d\theta^2 + \sin^2\theta\,d\phi^2\right),
\label{eq:1}
\end{equation}
where 
\begin{equation}
f(r) = 1 - \frac{2M}{\sqrt{r^2 + a^2}} + \frac{\zeta}{r} \log\frac{r}{|\zeta|}.
\end{equation}
Here, $M$ denotes the black hole mass, $a$ is a parameter characterizing the magnetic charge, and $\zeta$ is the PFDM parameter, in the limit where $a \to 0$ and $\zeta \to 0$, it degenerates to the Schwarzschild black hole.

According to the strong-field limit theory of gravitational lensing \cite{Bozza:2002zj}, an extremal black hole is characterized by the coincidence of its event horizon and the innermost unstable photon orbit(photon sphere). In this study, the critical curve corresponding to extremal black holes in the parameter space $(a/M,\zeta/M)$ is determined by simultaneously solving $f(r)=0$ (the horizon condition) and $f'(r)=0$ (the photon-sphere condition). This curve delineates the physically admissible region where black hole solutions exist, as shown in Fig.1, where $a/M$ is the dimensionless magnetic charge parameter and $\zeta/M$ is the dimensionless PFDM parameter. The solid blue curve represents the trajectory of extremal black holes, corresponding to the coincidence of the event horizon and the inner horizon. The gray shaded region indicates the parameter range where regular black holes with two horizons exist, while the white region denotes the parameter space where no black hole solution (i.e., no physical horizon) is present. Furthermore, for the convenience of subsequent numerical analysis, the parameters $a$ and $\zeta$ appearing throughout this paper are dimensionless, i.e., $a \equiv a/M$ and $\zeta \equiv \zeta/M$.

To constrain the parameters of the charged-PFDM black hole using the shadow data observed by the EHT, we follow the methodology outlined in \cite{EventHorizonTelescope:2022wkp, Perlick:2021aok}. For a static spherically symmetric black hole, the radius of the photon ring $r_{\text{ph}}$ is determined by:
\begin{equation}
	   r f'(r) = 2 f(r), 
\end{equation}
for an asymptotically flat metric background, the shadow radius $R_s$ of  black hole as observed at infinity is defined by \cite{EventHorizonTelescope:2022wkp}
\begin{equation}
R_s = \frac{r_{\mathrm{ph}}}{\sqrt{f(r_{\mathrm{ph}})}}.
\end{equation}
\begin{figure}[t]
\centering
\includegraphics[width=0.8\columnwidth]{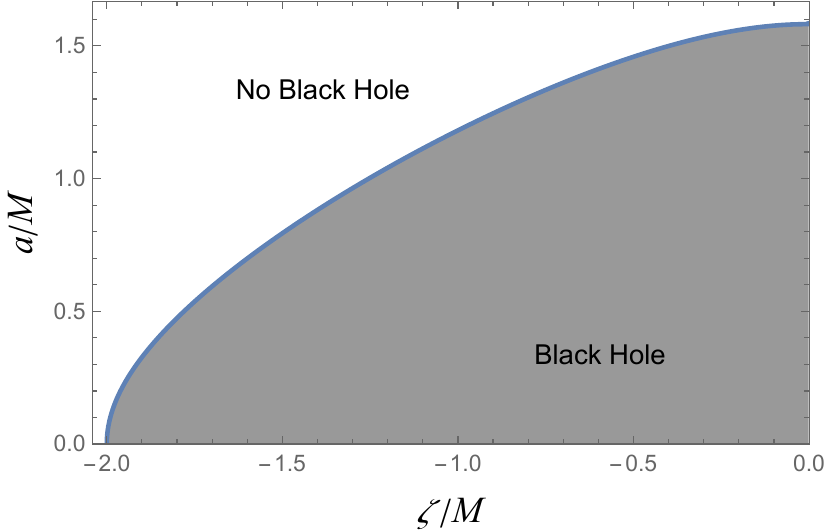}
\caption{\justifying The phase diagram for the parameter space $(a/M,\zeta/M)$ of charged-PFDM black hole}
\label{fig:1}
\end{figure}

In 2019, the Event Horizon Telescope (EHT) Collaboration successfully captured the first shadow image of the supermassive black hole M87* at the center of the elliptical galaxy M87 \cite{EventHorizonTelescope:2019dse, EventHorizonTelescope:2019ths, EventHorizonTelescope:2021dqv, EventHorizonTelescope:2019ggy}. The observations yielded an angular shadow diameter of $\theta_{\mathrm{sh}} = (42 \pm 3)\,\mu\mathrm{as}$, a distance of $D = (16.8 \pm 0.8)\,\mathrm{Mpc}$, and a mass estimate of  $(6.5 \pm 0.7) \times 10^9 \, M_\odot$. This landmark achievement provided the first strong constraints on theoretical models from event-horizon scales \cite{Vachher:2024ldc}. Subsequently, in 2022, the EHT reported the shadow observation of Sgr A*, the black hole at the Galactic Center \cite{EventHorizonTelescope:2022wkp, EventHorizonTelescope:2022urf, EventHorizonTelescope:2022xqj}. The measured angular diameter is $\theta_{\mathrm{sh}} = (48.7 \pm 7)\,\mu\mathrm{as}$, the distance is $D = 8\,\mathrm{kpc}$, and the mass is about$(4.0_{-0.6}^{+1.1}) \times 10^6 \, M_\odot$. These high-precision data establish a solid foundation for testing various black hole models \cite{Saurabh:2020zqg, Vagnozzi:2022moj}.

\begin{figure}[h]
\centering
\includegraphics[width=0.8\columnwidth]{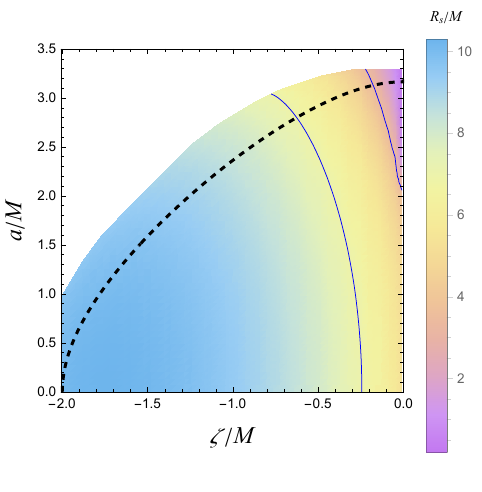}
\caption{\justifying EHT-constrained parameter space of M87* candidate charged black holes in PFDM background. The horizontal axis is PFDM parameter $\zeta/M$, left vertical axis is magnetic charge parameter $a/M$, and the right color bar links to $R_s/M$. The black dashed line marks extreme black holes (black holes exist below it); the area between two curves is constrained by EHT observations of M87*.}
\label{fig:2}
\end{figure}

Here, we utilize the shadow measurement of M87*, with a reported radius of $3.430 \leq R_s \leq 6.963 \; (2\sigma)$ \cite{EventHorizonTelescope:2022wkp}, to constrain the parameter space of the charged-PFDM black hole. Fig 2 illustrates the parameter space constrained by EHT observations when the black hole is considered a candidate for M87*. The allowed region lies between two characteristic curves and below the critical line for extremal black holes (black dashed line). The analysis shows that when the magnetic charge parameter $a/M$ is small, the effective range of the dark matter parameter $\zeta/M$ is approximately between $-0.8$ and $0$. As a/M increases, this range gradually extends toward more negative values, eventually covering $-1.7$ to $-0.8$. This work focuses on the constraints imposed by EHT observations in a dark matter dominated background. Synthesizing the above results, the parameter ranges adopted in the subsequent analysis are $0 < a \leq 1$ and $-0.3\leq \zeta<0$.

While Sgr A* data ostensibly give tighter parameter constraints \cite{EventHorizonTelescope:2021dqv}, they lead to $\zeta \approx 0$ in practice, effectively nullifying the dark matter contribution. This is likely due to the intricate Galactic Center environment, where interstellar scattering, magnetic interference, and unsubtracted foreground radiation amplify the sensitivity of the shadow measurement. By contrast, M87*—situated in a tranquil elliptical galaxy—provides cleaner observations that better capture the true coupling of general relativity with dark matter. Consequently, we employ the $(2\sigma)$ constraints from M87* to maintain physical consistency and reliability in our study. In this work, the above constraint has been consistently adopted throughout our analysis.

\section{Thin accretion disk around considered black hole}
\label{sec:3}

This section investigates the dynamical behavior of the thin accretion disk. We begin by deriving the effective potential from the equations of motion, which provides key insights into the physical factors governing the particle's trajectory. Our analysis is based on the Lagrangian $\mathcal{L}$ for a point particle moving in the spacetime described by:

 \begin{equation}
   \mathcal{L} = \frac{1}{2} g_{\mu\nu} \frac{dx^\mu}{ds} \frac{dx^\nu}{ds} = \frac{1}{2} \varepsilon,
   \label{eq:5}
\end{equation}

The Lagrangian presented in Eq.\eqref{eq:5} describes massive particles for $\varepsilon=1$ and photons for $\varepsilon=0$. Our analysis is confined to equatorial orbits with $\theta=\frac{\pi}{2}$.The conserved energy E and angular momentum L are obtained as:

\begin{equation}
E = -g_{tt} \frac{dt}{ds} 
= \left(1 - \frac{2M}{\sqrt{r^2 + a^2}} + \frac{\zeta}{r} \log\frac{r}{|\zeta|}\right) \frac{dt}{ds},
\label{eq:6}
\end{equation}

\begin{equation}
	L = g_{\phi\phi} \frac{d\phi}{ds} 
	= r^2 \frac{d\phi}{ds}.
	\label{eq:7}
\end{equation}

The geodesic equations for a massive particle in the spacetime \eqref{eq:1} take the following forms:

\begin{align}
	\left(\frac{dr}{ds}\right)^2 &= E^2 - \left(1 - \frac{2M}{\sqrt{r^2 + a^2}} + \frac{\zeta}{r} \log\frac{r}{|\zeta|}\right) \left(1 + \frac{L^2}{r^2}\right), \label{eq:8} \\
	\left(\frac{dr}{d\phi}\right)^2 &= \frac{r^4}{L^2} \left[ E^2 - \left(1 - \frac{2M}{\sqrt{r^2 + a^2}} + \frac{\zeta}{r} \log\frac{r}{|\zeta|}\right) \left(1 + \frac{L^2}{r^2}\right) \right], \label{eq:9} \\
	\begin{split}
		\left(\frac{dr}{dt}\right)^2 &= \frac{1}{E^2} \left(1 - \frac{2M}{\sqrt{r^2 + a^2}} + \frac{\zeta}{r} \log\frac{r}{|\zeta|}\right)^2 \\
		&\quad \times \left[ E^2 - \left(1 - \frac{2M}{\sqrt{r^2 + a^2}} + \frac{\zeta}{r} \log\frac{r}{|\zeta|}\right) \left(1 + \frac{L^2}{r^2}\right) \right]. \label{eq:10}
	\end{split}
\end{align}

The system's dynamics are completely governed by Eqs.\eqref{eq:8}\eqref{eq:9}\eqref{eq:10}, with the effective gravitational potential given directly by Eq.\eqref{eq:8}:
\begin{equation}
 V_{\text{eff}}(r)=-g_{tt}(1+\frac{L^2}{r^2})=(1 - \frac{2M}{\sqrt{r^2 + a^2}} + \frac{\zeta}{r} \log\frac{r}{|\zeta|})(1+\frac{L^2}{r^2}).
\label{eq:11}
\end{equation}
When a test particle travels along a circular orbit, its radius must satisfy the following constraints:

\begin{equation}
V_{\text{eff}} = E^2, \quad V_{\text{eff},r} = 0.
\label{eq:12}
\end{equation}

Through the solution of the above two equations, the explicit dependence of $\Omega$, $E$, $L$ on the circular orbit radius $r$ can be obtained:

\begin{equation}
	\begin{aligned}
		\Omega &= \frac{d\phi}{dt} = \sqrt{\frac{-g_{tt,r}}{g_{\phi\phi,r}}} 
		= \sqrt{\frac{M}{(r^2 + a^2)^{3/2}} + \frac{\zeta\left(1 - \log\frac{r}{|\zeta|}\right)}{2r^3}}, 
	\end{aligned}
	\label{eq:13}
\end{equation}

\begin{equation}
	\begin{aligned}
		E &= \frac{-g_{tt}}{\sqrt{-g_{tt} - g_{\varphi\varphi}\Omega^2}}  \\
		&= \frac{1 - \dfrac{2M}{\sqrt{r^2 + a^2}} + \dfrac{\zeta}{r}\log\dfrac{r}{|\zeta|}}
		{\sqrt{1 - \dfrac{2M}{\sqrt{r^2 + a^2}} - \dfrac{Mr^2}{(r^2 + a^2)^{3/2}} + \dfrac{\zeta}{2r}\left(3\log\dfrac{r}{|\zeta|} - 1\right)}},
	\end{aligned}
	\label{eq:14}
\end{equation}

\begin{equation}
	\begin{aligned}
		L &= \frac{g_{\varphi\varphi}}{\sqrt{-g_{tt} - g_{\varphi\varphi}\Omega^2}} \\
		&= \sqrt{\frac{
				\dfrac{2Mr^4}{(r^2 + a^2)^{3/2}} + \zeta r\left(1 - \log\dfrac{r}{|\zeta|}\right)
			}{
				2 - \dfrac{4M}{\sqrt{r^2 + a^2}} - \dfrac{2Mr^2}{(r^2 + a^2)^{3/2}} + \dfrac{\zeta}{r}\left(3\log\dfrac{r}{|\zeta|} - 1\right)
		}},
	\end{aligned}
	\label{eq:15}
\end{equation}

In general, not all circular orbits are stable. An interesting critical circular orbit is the innermost stable circular orbit (ISCO), which is determined by the equations $V_{\text{eff}}=0$, $\quad V_{\text{eff},r} = 0$, $\quad V_{\text{eff},rr} = 0$. Therefore, the radius $r_{isco}$ of the ISCO satisfies the following equation:

\begin{equation}
r_{isco} = \frac{2r f''(r) f(r) - 4r f'(r)^2 + 6 f(r) f'(r)}{2r f(r) - r^2 f'(r)}.
	\label{eq:16}
\end{equation}

\begin{figure*}
	a) \includegraphics[width=8.1 cm]{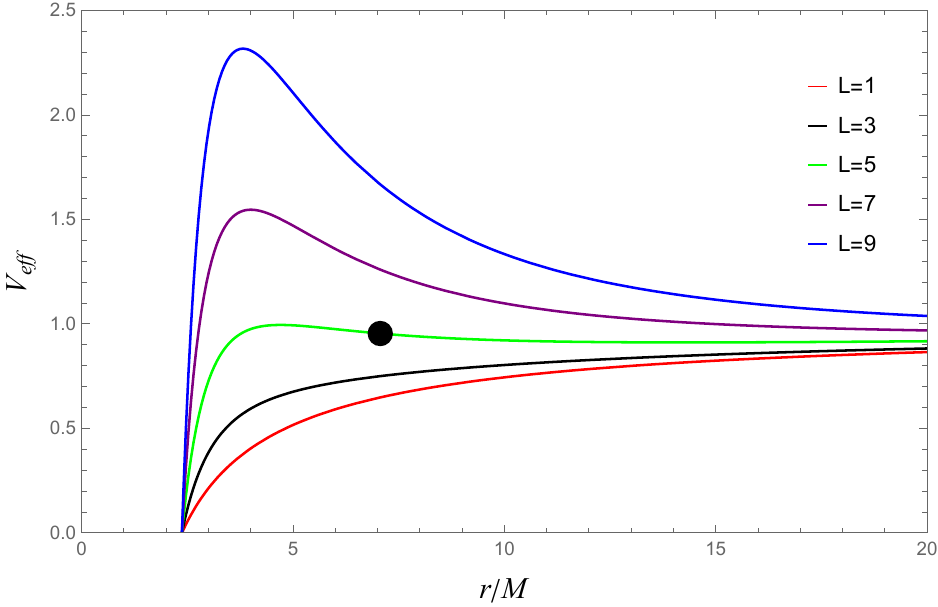}
	b) \includegraphics[width=8.1 cm]{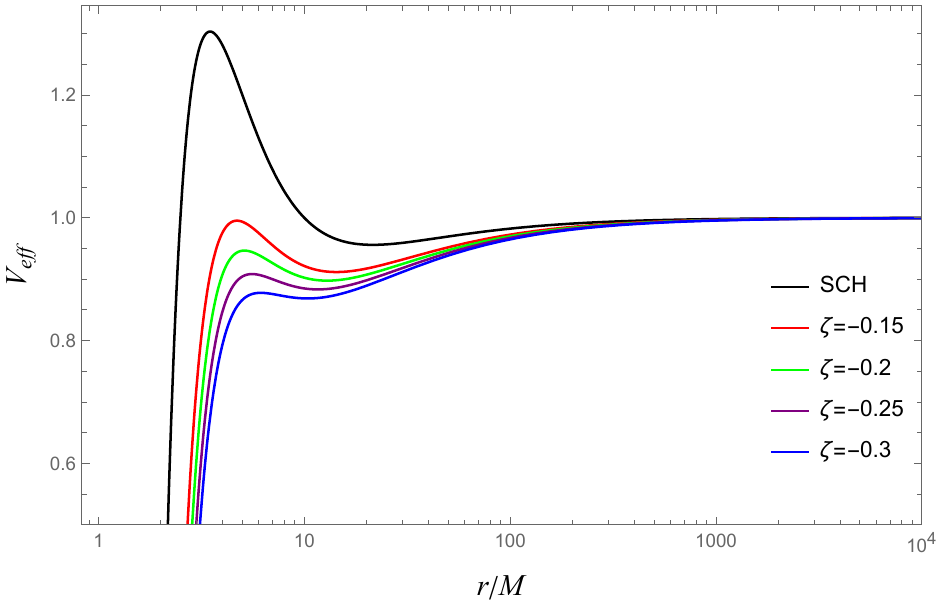}\\
	c) \includegraphics[width=8.1 cm]{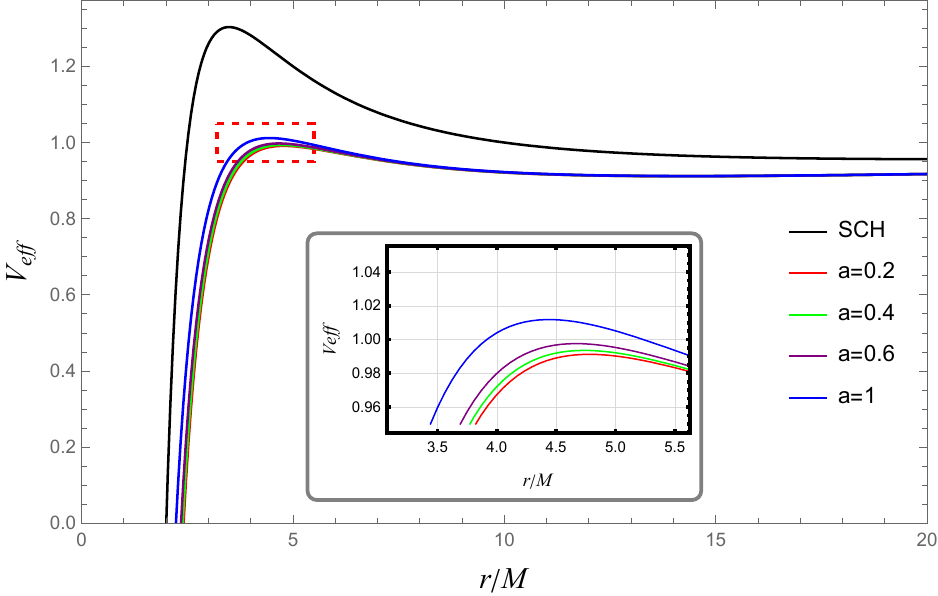}
	\caption{ The effective potential $V_{eff}$ depends on the radial coordinate $r$, and its variation characteristics under different parameter combinations are as follows (a) for $\zeta=-0.15$, $a=0.5$, with different values of $L$ (b) for $L=5$, $\zeta=-0.15$ with different values of $a$ (c) for $L=5$, $a=0.5$ with different values of $\zeta$.} \label{fig: 3}
\end{figure*}

Assuming the magnetic charge parameter $a=0.5$, PFDM parameter $\zeta=-0.15$, and black hole mass $M=1$, the ISCO radius $r_{isco}\approx 7.06414$ is obtained by solving Eq.\eqref{eq:16}. The radial profile of the effective potential $V_{\text{eff}}$ for a test particle is presented in Fig.3a reveals the first extremum occurring at $L=5$, with no additional extrema for $L<5$. The effective potential is further observed to increase with the angular momentum $L$. The ISCO, marked by a black dot in Fig.3a, is located at $r=7.06414$. In Fig.3b, a larger magnetic charge parameter $a$ leads to a higher effective potential, indicating that the trajectories of both unstable and stable circular orbits are situated farther from the central mass. As shown in Fig.3c, the effective potential decreases as the PFDM parameter $\zeta$ increases. Plotted on a logarithmic scale, a secondary minimum at larger $r$ becomes apparent; for particles with sufficiently large $L$,  $V_{\text{eff}}$ thus exhibits two extrema—a maximum (unstable circular orbit) and a minimum (stable circular orbit). Furthermore, comparison between Fig.3b and Fig.3c indicates that the effective potential of particles around the charged-PFDM black hole remains consistently lower than that in the Schwarzschild case.

Fig.4 shows the angular velocity $\Omega$ of particles on circular orbits as a function of orbital radius $r$. The angular velocity decreases monotonically with increasing $r$. The influence of the PFDM parameter $\zeta$ and the magnetic charge parameter $a$ on the orbital angular velocity is analyzed. The results indicate that $\Omega$ decreases with increasing values of $a$ or $\zeta$. However, outside the event horizon, it is consistently greater than that in the case of the Schwarzschild black hole. This implies that, at the same orbital radius, the gravitational field of the charged-PFDM black hole is stronger than that of a Schwarzschild black hole.

\begin{figure*}
	a)\includegraphics[width=8.1 cm]{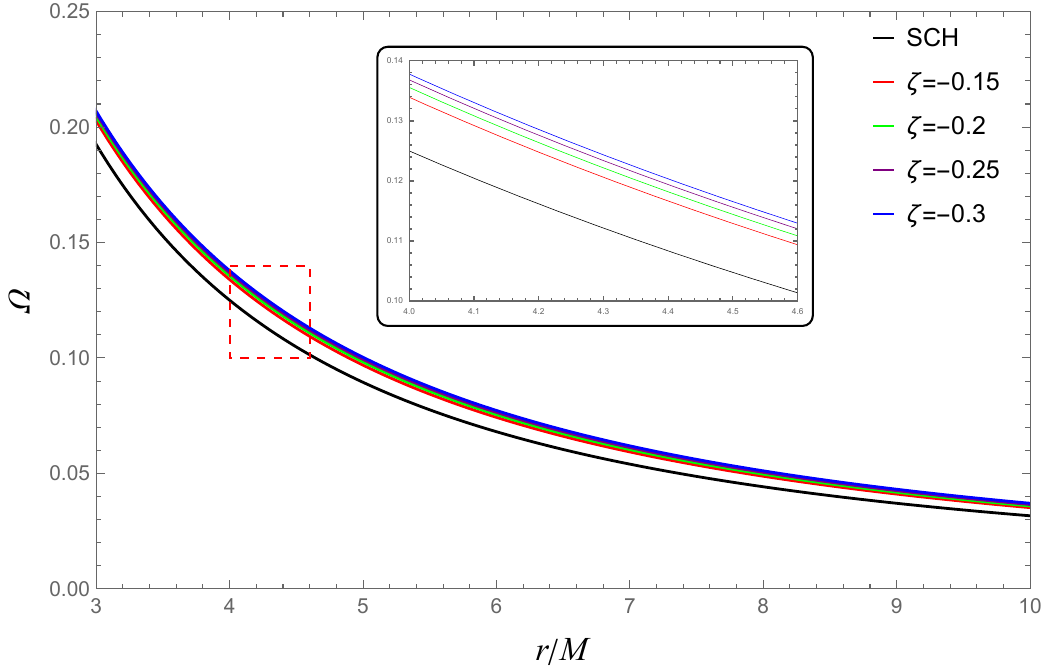}
	b)\includegraphics[width=8.1 cm]{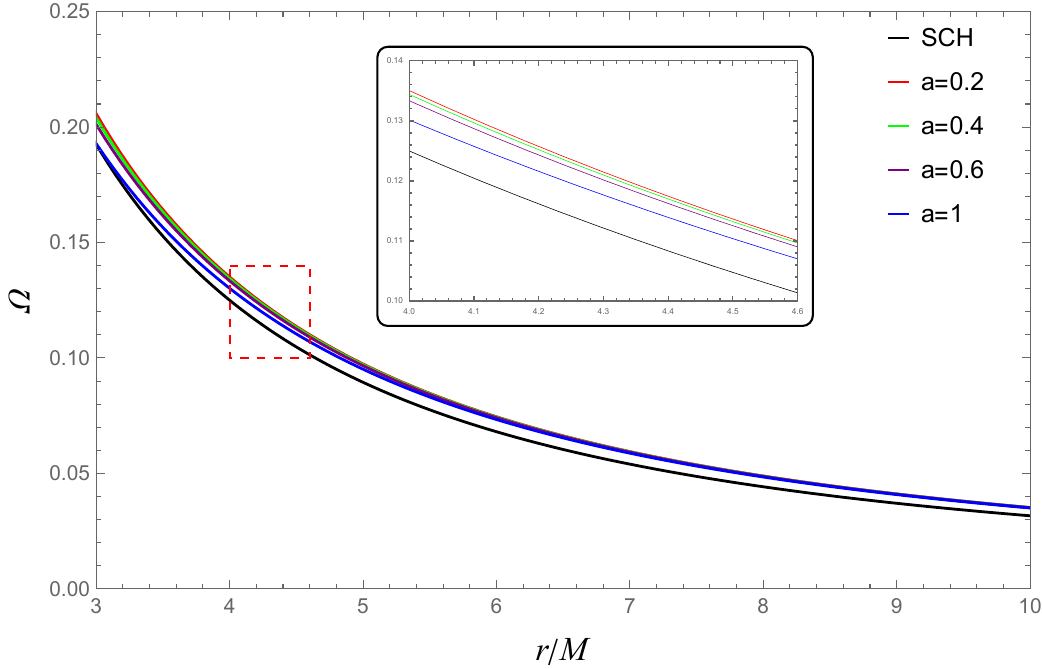}
	\caption{Angular velocity $\Omega$ versus $r$ (a) for $a=0.5$, varying $\zeta$, (b) for $\zeta=-0.15$, varying $a$.} \label{fig: 4}
\end{figure*}

The radial profiles of specific energy and specific angular momentum are displayed in Fig.5. From the left panel it is clearly seen that both specific energy and specific angular momentum decrease as the PFDM parameter $\zeta$ increases. The right panel shows that they also decline with rising magnetic charge $a$. Notably, at the same radius $r$, the angular momentum of the charged-PFDM black hole remains consistently higher than that of the Schwarzschild black hole, while its specific energy drops below the Schwarzschild value beyond a certain radial distance.

\begin{figure*}
	(a)\includegraphics[width=8.1 cm]{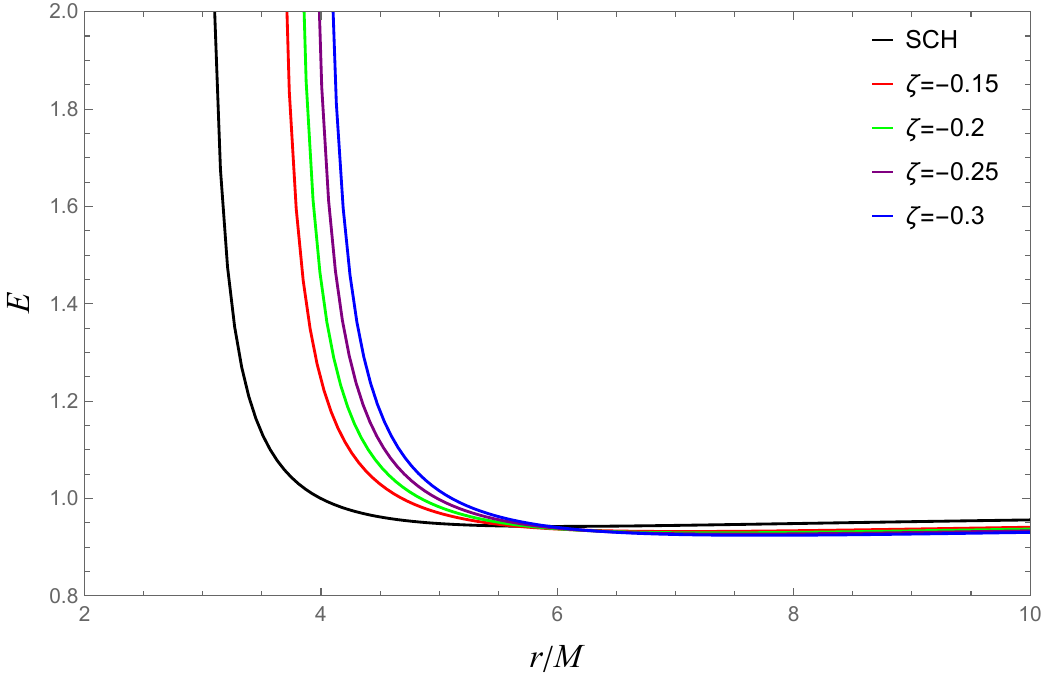}
	(b)\includegraphics[width=8.1 cm]{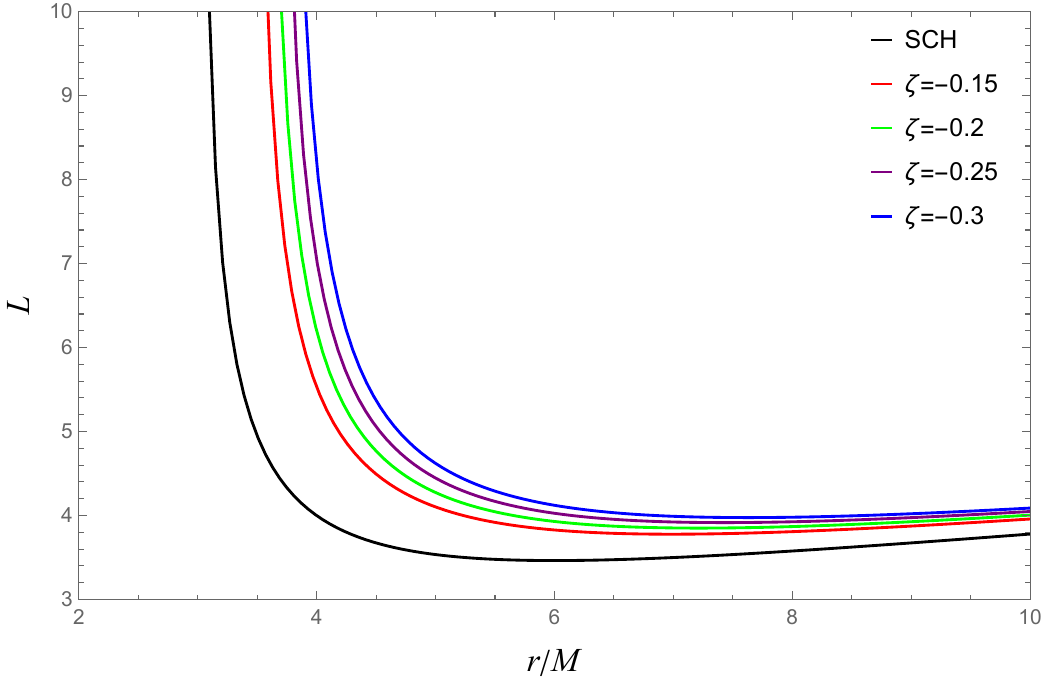}
	(c)\includegraphics[width=8.1 cm]{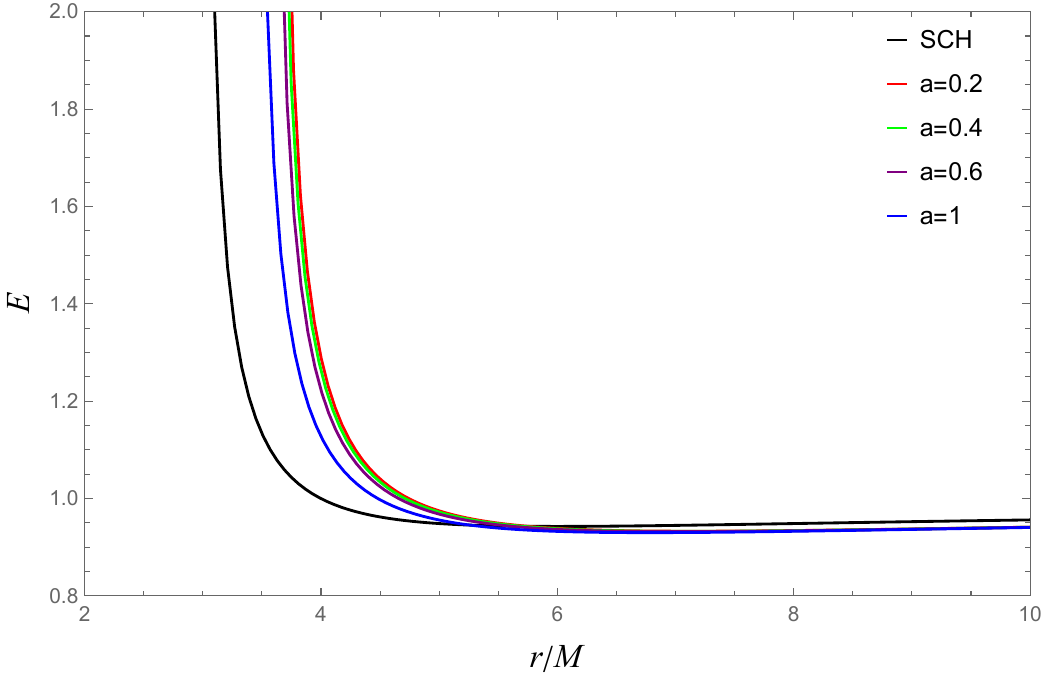}
	(d)\includegraphics[width=8.1 cm]{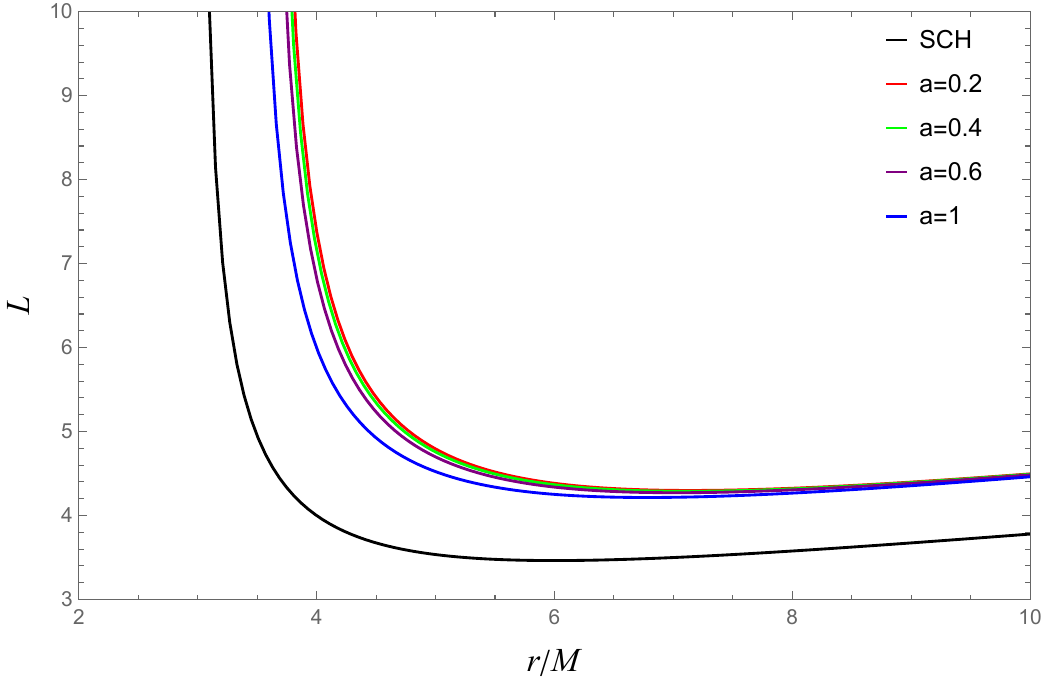}
	\caption{ The distributions of specific energy and specific angular momentum are shown as functions of the radial coordinate $r$. The left panel displays results for different PFDM parameter $\zeta$ values with $a = 0.5$ , while the right panel corresponds to different magnetic charge parameter a values with $\zeta=-0.15$.}. \label{fig: 5}
\end{figure*}

\section{Properties of thin accretion disk}
\label{sec:4}

In this section, we investigate the physical properties based on the Novikov-Thorne model of a thin accretion disk \cite{Novikov:1973kta}. Within a stationary, axisymmetric, and asymptotically flat spacetime background, the thickness of a thin accretion disk is typically negligible, meaning its vertical half-thickness H is always much smaller than its characteristic radial distance $r$ $(H \ll r)$. The thin disk is in a state of hydrodynamic and thermodynamic equilibrium. Pressure gradients and vertical entropy gradients within the accreting material are negligible, and the heat generated by viscous stresses is efficiently radiated away from the disk surfaces. This efficient radiative cooling prevents the disk from puffing up and maintains its slender vertical structure. The inner edge of the disk is defined by the ISCO within the gravitational potential of the central compact object, located at $r_{isco}$. Furthermore, the mass accretion rate $\dot{M}_0$ is assumed to be constant over time.

In the following subsection, we take the central black hole of M87* to be described by the charged-PFDM model and conduct a detailed investigation of its radiative energy flux, temperature profile, observed luminosity, and other key quantities associated with a thin accretion disk. Since the Novikov-Thorne model requires an asymptotically flat spacetime, we primarily focus on the case of a vanishing cosmological constant. The specific values of the physical constants and the adopted thin-disk parameters are summarized in \textbf{Table I}.

\subsection{Radiant energy flux}

In investigating the radiative flux from the outer region of the accretion disk within the equatorial plane, we employ the quantities $E$, $L$, and $\Omega$. The radiant energy flux across the disk can be determined using the following relation \cite{Novikov:1973kta, Page:1974he, Collodel:2021gxu}:

\begin{align}
	F(r) = -\frac{\dot{M}_0 \Omega_{,r}}{4\pi \sqrt{-g/g_{\theta\theta}} (E - \Omega L)^2} 
	\int_{r_{\mathrm{isco}}}^{r} (E - \Omega L) L_{,r} \, dr,
	\label{eq:17}
\end{align}

where $\dot{M}_0$ is the mass accretion rate. Fig.6 presents the profiles for the radiative energy flux $F(r)$ from accretion disks surrounding the charged-PFDM black hole for different values of $\zeta$ or $a$. For comparison, the flux distribution for a disk around a Schwarzschild black hole is also included. It is observed that the energy flux in the charged-PFDM geometry is consistently lower than that in the standard Schwarzschild case. Furthermore, the flux increases with larger values of the parameters $\zeta$ or $a$, indicating that the deviation from the standard Schwarzschild result diminishes as these black-hole parameters grow. In addition, the peak (maximum) of the flux shifts toward smaller radii as the black hole parameters increase. These characteristics provide potential avenues to constrain the parameters $\zeta$ and $a$ through future astronomical observations.

\begin{figure*}
	a)\includegraphics[width=8.1 cm]{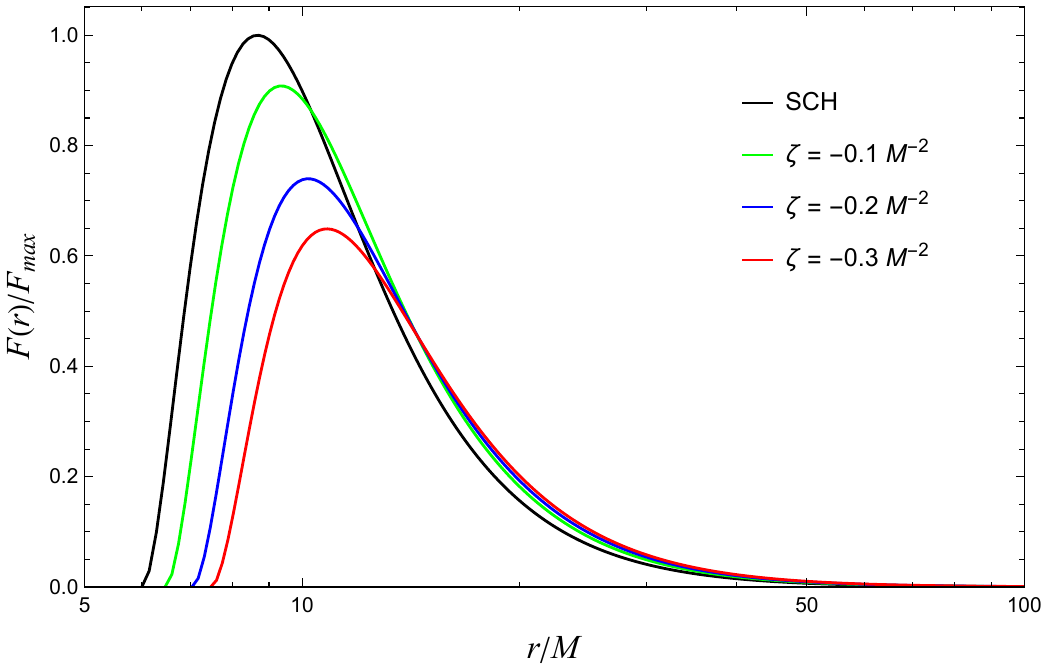}
	b)\includegraphics[width=8.1 cm]{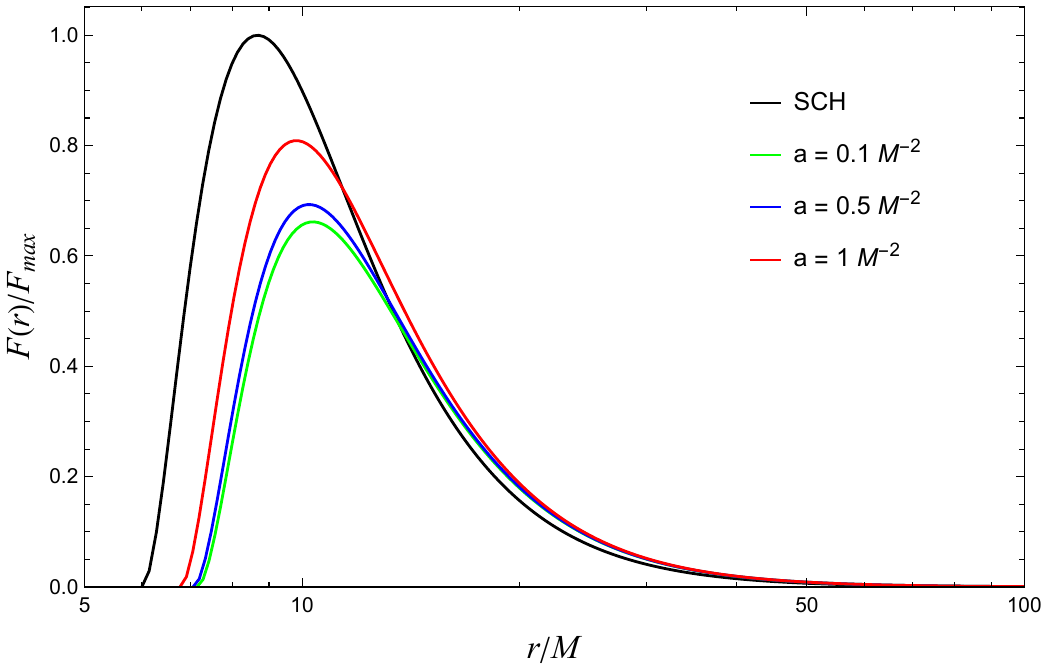}
	\caption{Radiative energy comparison between accretion disks around charged-PFDM and Schwarzschild black holes of equal total mass (a) different $\zeta$ values with fixed $a=0.5$ (b) different $a$ values with fixed $\zeta=-0.15$. Flux values are normalized to the maximum flux of the Schwarzschild case, $F_{\max} = 1.37 \times 10^{-5} \, \dot{M}_0/M^2$.} \label{fig: 9}
\end{figure*}

\subsection{Radiant temperature}

Within the Novikov-Thorne framework, the accreting matter maintains thermodynamic equilibrium. Consequently, the radiation emanating from the disk can be treated as perfect blackbody radiation. The disk's radiative temperature $T(r)$ is connected to its energy flux $F(r)$ through the Stefan-Boltzmann law:

\begin{equation}
	F(r) = \sigma_{\mathrm{SB}} T^4,
	\label{eq:18}
\end{equation}

where $\sigma_{\mathrm{SB}}$ is the Stefan-Boltzmann constant. This indicates that the radial profile of the temperature $T(r)$ follows a trend similar to that of the energy flux $F(r)$. As shown in Fig.7, the temperatures are normalized to the maximum value for a standard Schwarzschild black hole. In Fig.7a (with fixed $a=0.5$), the disk temperature increases for larger values of the PFDM parameter $\zeta$, and the temperature peak shifts toward the inner edge of the disk. Similarly, Fig.7b (with fixed $\zeta=-0.15$) shows that the temperature rises with increasing magnetic charge parameter $a$, accompanied by an inward shift of the peak. Overall, the accretion disk around the charged-PFDM BH is cooler to the disk around a Schwarzschild BH.

\begin{figure*}
	a)\includegraphics[width=8.1 cm]{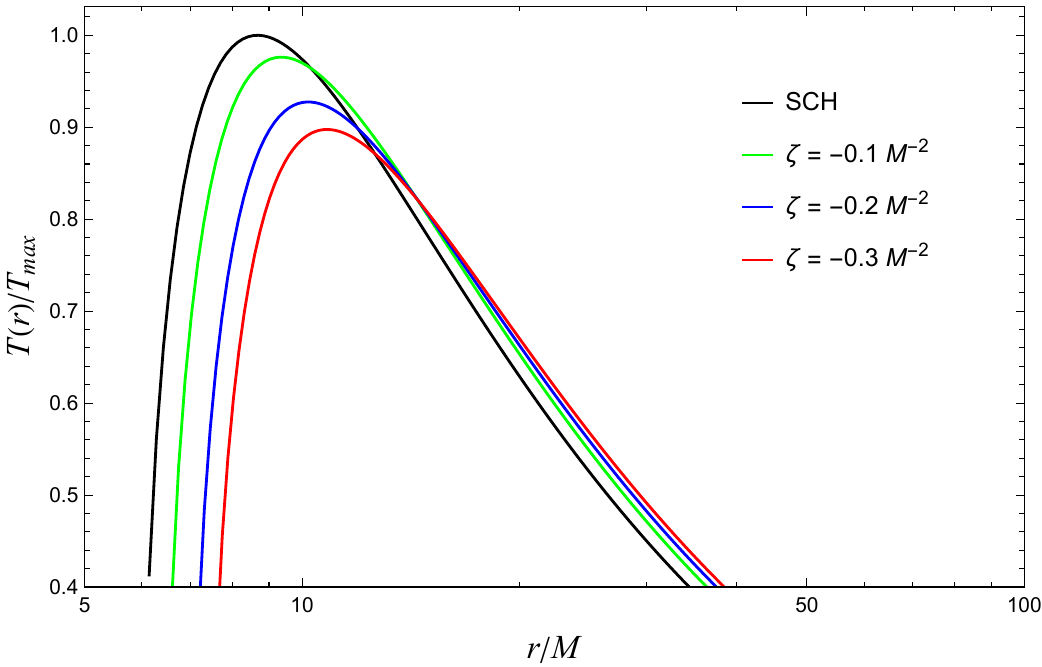}
	b)\includegraphics[width=8.1 cm]{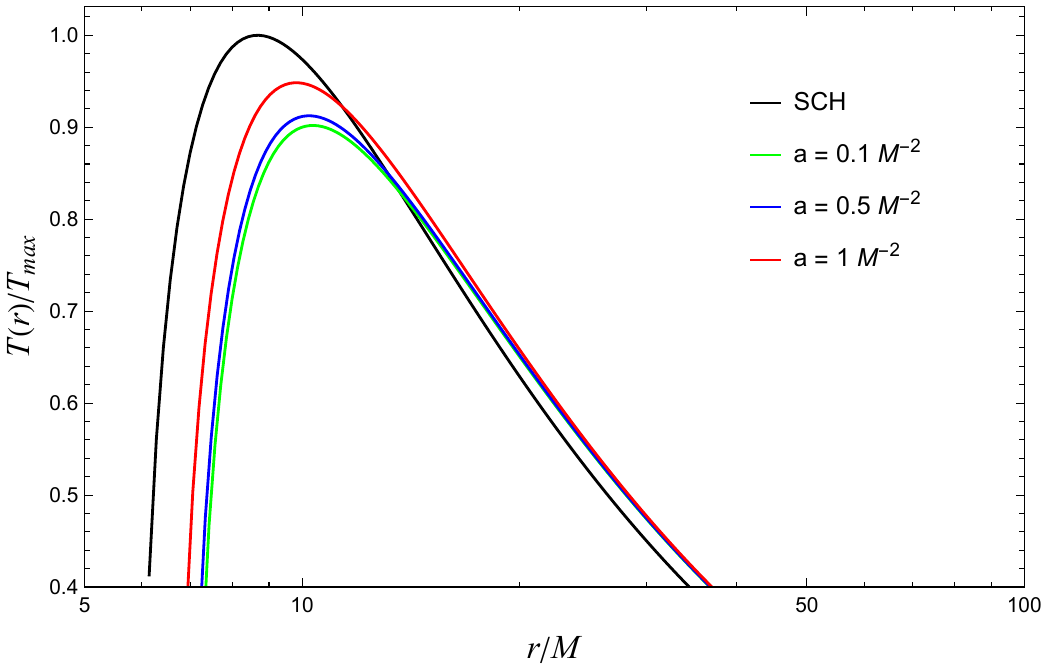}
	\caption{The disk temperature for charged-PFDM and Schwarzschild black holes of equal total mass (a) different $\zeta$ values with fixed $a=0.5$ (b) different $a$ values with fixed $\zeta=-0.15$. Temperature values are normalized to the maximum temperature of the Schwarzschild case, $T_{\max} = 0.7015 \, (\dot{M}_0/M^2)^{1/4}$.} \label{fig: 10}
\end{figure*}

\subsection{Observed luminosity}

Under the assumption that disk-emitted radiation has a blackbody spectrum, the disk’s monochromatic luminosity $L(\nu)$ can be calculated by:

\begin{align}
	L(\nu) &= 4\pi d^2 I(\nu) \nonumber \\
	&= 8\pi h \cos\gamma \int_{r_i}^{r_f} \int_0^{2\pi} 
	\frac{\nu_e^3 r}{e^{h\nu_e/(k_B T)} - 1} \, dr \, d\phi,
	\label{eq:19}
\end{align}

where $d$ is the distance to the disk center, $\nu$ is the frequency of emitted photons, $I(\nu)$ stands for the thermal energy flux corresponding to a specific frequency, $h$ is the Planck constant and $k_B$ is the Boltzmann constant. $\gamma$ denotes the disk’s inclination angle which we set it to zero. $r_i$ and $r_f$ denote the positions of the inner and outer edges of the disk, respectively. To ensure that the flux over the disk surface vanishes at $r_f \to \infty$ for any asymptotically flat geometry, we set $r_i =r_{isco}$ and take $r_f \to \infty$. Also, $\nu_e= \nu(1+z)$ corresponds to the emitted photon frequency, with $z$ being the redshift factor. The redshift factor $z$ is computed without accounting for light bending effects\cite{Luminet:1979nyg, Bhattacharyya:2000kt, Salahshoor:2018plr}:

\begin{equation}
1 + z = \frac{1 + \Omega r \sin\varphi \sin\gamma}{\sqrt{-g_{tt} - 2\Omega g_{t\varphi} - \Omega^2 g_{\varphi\varphi}}},
	\label{eq:20}
\end{equation}

Employing Eqs.\eqref{eq:19} and \eqref{eq:20}, we calculate the spectral energy distribution (SED) of the thin accretion disk around the black hole, the results are presented in Fig.8. This figure compares the radiation spectra of the charged-PFDM black hole with that of the Schwarzschild black hole. For all parameter choices, the SED of the charged-PFDM black hole lies systematically above the Schwarzschild case, in agreement with the higher radiative efficiency $\eta^*$ shown in \textbf{Table II}, which implies a higher total luminosity for the same accretion rate. Specifically, in Fig.8a (fixed $a=0.5$), a more negative PFDM parameter $\zeta$ leads to a larger ISCO radius (see \textbf{Table II}), an increase in radiative efficiency and total luminosity, and consequently an upward shift of the entire SED. In Fig.8b (fixed $\zeta=-0.15$), as the magnetic charge parameter $a$ increases, the deepening gravitational potential raises the radiative efficiency and simultaneously shifts the spectral peak toward higher frequencies (blueshift). These characteristics suggest that observations of the continuum spectrum of active galactic nuclei could provide important observational constraints on the magnetic charge a and the dark matter parameter $\zeta$ surrounding black holes.

\begin{figure*}
	a)\includegraphics[width=8.1 cm]{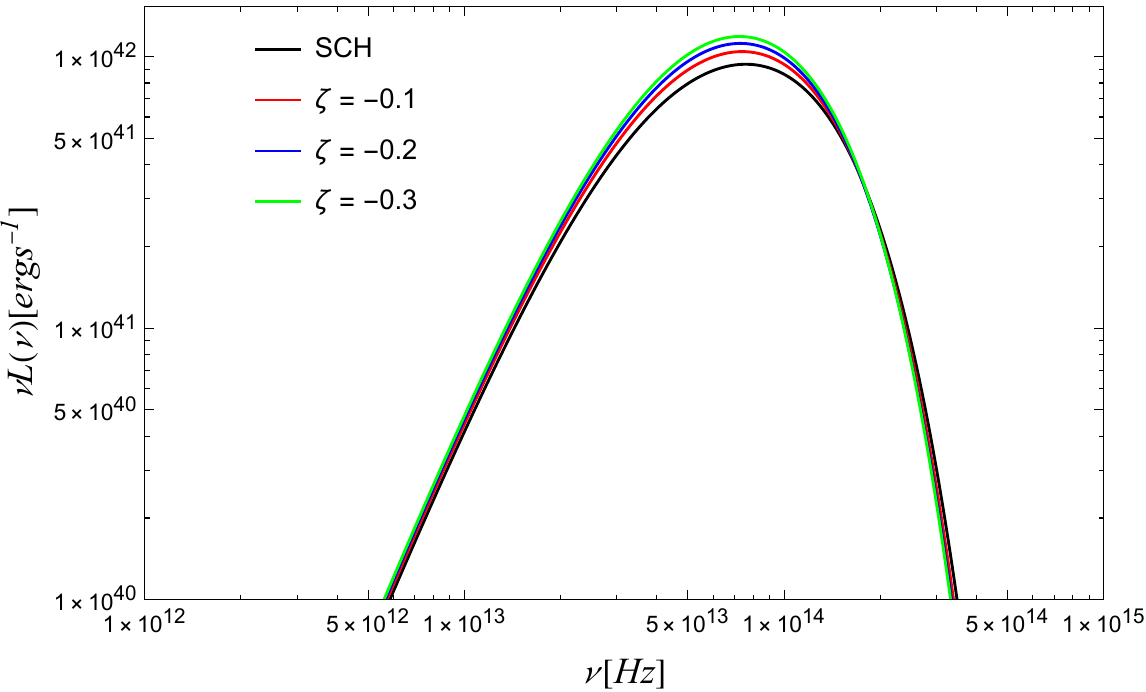}
	b)\includegraphics[width=8.1 cm]{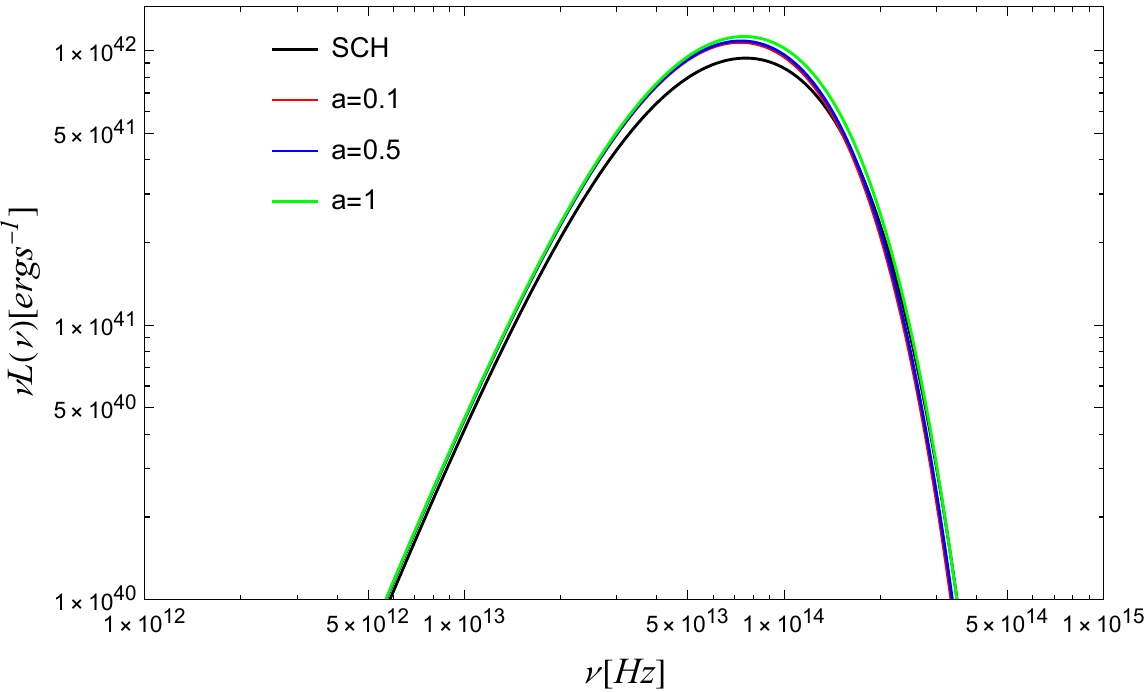}
	\caption{The disk spectra for charged-PFDM and Schwarzschild black holes of same total mass (a) different $\zeta$ values with fixed $a=0.5$ (b) different $a$ values with fixed $\zeta=-0.15$.} \label{fig: 11}
\end{figure*}

\subsection{ Radiative efficiency}

A key physical quantity associated with mass accretion is the radiation efficiency of the black hole. This efficiency is defined as the ratio of the photon energy radiated to infinity from the disk surface to the rate at which rest mass energy is supplied to the central compact object, with both quantities measured at infinity \cite{Novikov:1973kta, Page:1974he}. Considering a test particle of unit mass falling from infinity down to the ISCO, and assuming that all the lost gravitational energy is converted into radiation that reaches infinity, the radiation efficiency $\eta^*$ is given by:

\begin{equation}
	\eta^* = 1 - E_{isco},
	\label{eq:21}
\end{equation}

where $E_{isco}$ is the specific energy at the ISCO. \textbf{Table II} lists the ISCO radius and the mass to radiation conversion efficiency for the charged-PFDM black hole under different parameter choices. Interestingly, when $\zeta$ is fixed at $–0.15$, the ISCO radius decreases with increasing $a$, whereas the radiative efficiency rises. As seen in Figs.6 and 7, both the flux $F(r)$ and the temperature $T(r)$ increase with larger $a$. This indicates that, although a smaller ISCO reduces the emitting area, the enhancement in the local flux and temperature dominates the increase in efficiency.

Conversely, with $a$ fixed at $0.5$, a more negative $\zeta$ leads to a higher efficiency. In this case, even though the local $F(r)$ and $T(r)$ are lower, the larger ISCO radius – and consequently the greater overall emitting area – results in a higher total radiated power. On the whole, the variation of the ISCO radius (and thus the disk area) appears to influence the thin disk behavior more significantly; the wider spread of the curves in Figs.6 and 7 for fixed $a$ and different $\zeta$ reflects this point.

For comparison, a Schwarzschild black hole has an efficiency of about $6\%$ (whether or not photon capture is included) and an ISCO at $r_{isco}=6M$ \cite{Page:1974he}. Therefore, although the local $F(r)$ and $T(r)$ of the charged-PFDM black hole are lower than those of the Schwarzschild case, its larger ISCO radius and a higher radiative efficiency and, for a fixed accretion rate, a higher total observed luminosity.

\begin{table}[htbp]
	\centering
	\footnotesize 
	\setlength{\heavyrulewidth}{0.12ex} 
	\setlength{\lightrulewidth}{0.08ex} 
	\setlength{\tabcolsep}{3pt}
	\resizebox{\columnwidth}{!}{
		\begin{tabular}{cc}
			\toprule 
			parameters & values  \\
			\midrule %
			$G$  & $6.67430 \times 10^{-11} \, \mathrm{m}^3 \, \mathrm{kg}^{-1} \, \mathrm{s}^{-2}$   \\
			$c$  & $2.99792458 \times 10^8 \, \mathrm{m/s}$   \\
			$h$  & $6.62607015 \times 10^{-27} \, \mathrm{erg\,s}$   \\
			$k_B$  & $1.380649 \times 10^{-16} \, \mathrm{erg\,K^{-1}}$   \\
			$M_\odot$  & $1.989 \times 10^{33} \, \mathrm{g}$   \\
			$1yr$  & $3.156 \times 10^{7} \, \mathrm{s}$  \\
			$\sigma_{\mathrm{SB}}$  & $5.670374419 \times 10^{-5} \, \mathrm{erg\,s^{-1}\,cm^{-2}\,K^{-4}}$  \\
			$M (M87^*)$  & $6.5 \times 10^{9} \, M_{\odot}$  \\
			$\dot{M}_0 (M87^*)$  & $3 \times 10^{-4} \, M_{\odot} \, \mathrm{yr^{-1}}$  \\
			
			\bottomrule 
		\end{tabular}
	}
	\caption{\justifying Physical constants and M87* baseline parameters adopted in this study} 
	\label{tab:1}
\end{table}

\begin{table*}[htbp]
	\centering
	\footnotesize
	\setlength{\heavyrulewidth}{0.12ex} 
	\setlength{\lightrulewidth}{0.08ex} 
	\setlength{\tabcolsep}{20pt} 
	\resizebox{\columnwidth}{!}{
		\begin{tabular}{cccc}
			\toprule
			$\zeta$ & $a$ & $r_{isco}$ & $\eta^*$  \\
			\midrule %
			$-0.1$ & $0.5$ & $6.7768$ & $0.0647$   \\
			$-0.2$ & $0.5$ & $7.3089$ & $0.0703$   \\
			$-0.3$ & $0.5$ & $7.7138$ & $0.0754$      \\
			$-0.15$ & $0.1$ & $7.1531$ & $0.0669$     \\
			$-0.15$ & $0.5$ & $7.0641$ & $0.0676$   \\
			$-0.15$ & $1$ & $6.7753$ & $0.0699$   \\
			
			\bottomrule
		\end{tabular}
	}
	\caption{The ISCO radius $r_{isco}$ and radiative efficiency $\eta^*$ for different values of $\zeta$ or $a$.}
	\label{tab:2}
\end{table*}

\section{Dynamical framework of accretion}
\label{sec:5}
In realistic astrophysical environments, black hole accretion predominantly occurs in a diffuse, quasi-spherical manner, in contrast to the high angular momentum configurations required for thin accretion disks \cite{Yuan:2014gma}. The PFDM model, as an effective theory capable of explaining galactic rotation curves, incorporates a perfect fluid description that naturally accommodates a hydrodynamic treatment of mass distribution and motion. This framework is particularly well-suited for investigating large-scale accretion processes involving low-angular-momentum material. In this section, we consider the simplest scenario of steady state, spherically symmetric accretion onto a black hole, where dark matter is modeled as an ideal fluid. The fluid is assumed to be sufficiently light such that its own gravitational field can be neglected. The steady-state assumption implies that the black hole mass grows slowly enough that the fluid distribution has time to adjust to the evolving black hole metric over the relevant spacetime scales.

\subsection{Basic constants}

 we follow the basic framework of accretion presented by Babichev and Shahzad et al.\cite{Babichev:2005py, Babichev:2013vji,Shahzad:2024ljt}. We consider the simplest scenario of steady state, spherically symmetric accretion onto a black hole, where dark matter is modeled as an ideal fluid. The fluid is assumed to be sufficiently light such that its own gravitational field can be neglected. The steady-state assumption implies that the black hole mass grows slowly enough that the fluid distribution has time to adjust to the evolving black hole metric over the relevant spacetime scales. In this scenario, its energy–momentum tensor can be expressed as:

\begin{equation}
	T^{\mu\nu} = (\rho + p) u^\mu u^\nu - p g^{\mu\nu},
	\label{eq:22}
\end{equation}

where $\rho$ is the density, $p$ is the dark energy pressure,and $u^\mu = dx^\mu/ds$ is the radial 4-velocity component.

For forward fluid motion and the accretion process (inward flow), the conditions $u^t>0$ and $u^r<0$ must be satisfied, respectively.Restricting to accretion in the equatorial plane $\theta=\pi/2$, the four-velocity takes the form $(u^t,u^r,0,0)$. Imposing the normalization condition $(u^\mu u_\mu=1)$ then gives:

\begin{equation}
	u^t = \sqrt{\frac{1 - \frac{2M}{\sqrt{r^2 + a^2}} + \frac{\zeta}{r} \log\frac{r}{|\zeta|} + (u^r)^2}{1 - \frac{2M}{\sqrt{r^2 + a^2}} + \frac{\zeta}{r} \log\frac{r}{|\zeta|}}}.
	\label{eq:23}
\end{equation}

 Without loss of generality, we set $u^*=-u^r>0$. Considering the conservation of the energy-momentum tensor, integrating its radial component yields \cite{Shahzad:2024ljt}:

\begin{equation}
(\rho + p) u^* r^2 \sqrt{1 - \frac{2M}{\sqrt{r^2 + a^2}} + \frac{\zeta}{r} \log\frac{r}{|\zeta|} + (u^*)^2} = -C_1,
	\label{eq:24}
\end{equation}

where $C_1$ is the integration constant. By leveraging the coupling between the conservation principle and the four-velocity, the form described by the equation  $u_\mu T^{\mu\nu}_{;\nu}=0$, we obtain:

\begin{equation}
	(\rho + p) u^\mu_{;\nu} + u^\nu \rho_{,\nu} = 0,
	\label{eq:25}
\end{equation}

The particle number density $n$ is defined as:

\begin{equation}
	\frac{d\rho}{\rho + p} = \frac{dn}{n}.
	\label{eq:26}
\end{equation}

For the accretion fluid, we adopt an isothermal equation of state:

\begin{equation}
	p = k \rho,
	\label{eq:27}
\end{equation}

inserting into Eq.\eqref{eq:27} and performing the integration gives:

\begin{equation}
	n \propto \rho^{1/(1+k)},
	\label{eq:28}
\end{equation}

setting the proportionality constant to $1$, we obtain:

\begin{equation}
		n = \rho^{1/(1+k)},
	\label{eq:29}
\end{equation}

The conservation law $\nabla_\mu (n u^\mu) = 0$, when applied to a steady, spherically symmetric flow, reduces to:

\begin{equation}
    \frac{1}{r^2} \frac{d}{dr} (n u^r) = 0,
	\label{eq:30}
\end{equation}

By inserting Eq.\eqref{eq:29} into Eq.\eqref{eq:30} and then integrating, we obtain:

\begin{equation}
	u^* r^2 \rho^{1/(1+k)} = C_2,
	\label{eq:31}
\end{equation}

where $C_2$ denotes the integration constant. Given that $u^*>0$, it follows that $C_2>0$. Combining Eqs.\eqref{eq:24} and \eqref{eq:31} yields:

\begin{equation}
	\rho^{k/(1+k)} \sqrt{1 - \frac{2M}{\sqrt{r^2 + a^2}} + \frac{\zeta}{r} \log\frac{r}{|\zeta|} + (u^*)^2}  = C_3,
	\label{eq:32}
\end{equation}

we assume that the fluid is at rest at spatial infinity $u_\infty = 0$ with its density normalized to unity $\rho_\infty = 1$, so the $C_3 = \rho_\infty^{k/(1+k)} \sqrt{f_\infty + u_\infty} = 1$.

According to standard Bondi accretion theory, a critical point (sonic point) exists in spherically symmetric steady-state accretion for an isothermal fluid \cite{Bondi:1952ni}:

\begin{equation}
	C_s^2 = k.
	\label{eq:33}
\end{equation}

we consider relativistic accretion with the equation-of-state parameter taken as $k=0.5$. The critical point equation is given by:

\begin{equation}
	u_c^2 = \frac{k}{1-k} f(r_c),
	\label{eq:34}
\end{equation}
and
\begin{equation}
	\quad \frac{f'(r_c)}{f(r_c)} = \frac{4k}{(1-k) r_c}.
	\label{eq:35}
\end{equation}

Here, $r_c$ denotes the critical radius. By simultaneously solving Eqs.\eqref{eq:31}, \eqref{eq:33}, \eqref{eq:34} and \eqref{eq:35} for the critical point equation:

\begin{equation}
	C_2 = \frac{r_c^2}{2\sqrt{f_c}}.
	\label{eq:36}
\end{equation}

\textbf{Table III} lists $C_2$ under different BH parameters. For fixed $a=0.5$, $C_2$ rises with decreasing $\zeta$; for fixed $\zeta=-0.15$, it rises with increasing $a$. All values remain near $7.04$ (relative spread $<1\%$). we therefore fix $C_2$ at the mean value around $7.04$, i.e.

\begin{equation}
	\bar{C}_2 = 7.05.
	\label{eq:37}
\end{equation}

\begin{table}[htbp]
	\centering
	\footnotesize 
	\setlength{\heavyrulewidth}{0.12ex} 
	\setlength{\lightrulewidth}{0.08ex} 
	\setlength{\tabcolsep}{12pt}
	\resizebox{\columnwidth}{!}{
		\begin{tabular}{ccccc}
			\toprule 
			$\zeta$ & $a$ & $r_c$ & $f_c$ & $C_2$   \\
			\midrule %
			$-0.15$  & $0.5$ & $2.5123$ & $0.1985$ & $7.0425$ \\
			$-0.20$  & $0.5$ & $2.5186$ & $0.1993$ & $7.0626$ \\
			$-0.25$  & $0.5$ & $2.5250$ & $0.2001$ & $7.0829$ \\
			$-0.3$   & $0.5$ & $2.5314$ & $0.2009$ & $7.1034$ \\
			$-0.15$  & $0.2$ & $2.5112$ & $0.1982$ & $7.0376$ \\
			$-0.15$  & $0.4$ & $2.5118$ & $0.1984$ & $7.0400$ \\
			$-0.15$  & $0.6$ & $2.5128$ & $0.1987$ & $7.0451$ \\
			$-0.15$  & $1$   & $2.5148$ & $0.1992$ & $7.0553$ \\
			$0$      & $0$   &$2.5$     &$0.2$     & $6.9877$ \\
			
			\bottomrule 
		\end{tabular}
	}
	\caption{\justifying Critical-point parameters and the corresponding values of the integration constant $C_2$ for different combinations of BH parameters. The case $\zeta=0$ and $a=0$ corresponds to the Schwarzschild black hole.} 
	\label{tab:3}
\end{table}

\subsection{Dynamical parameters}

By simultaneously considering Eqs.\eqref{eq:31}, \eqref{eq:32} and \eqref{eq:37}, we derive:

\begin{equation}
	u^* = \frac{r^2 \pm \sqrt{r^4 - 4\bar{C}_2 \left(1 - \frac{2M}{\sqrt{r^2+a^2}} + \frac{\zeta}{r} \log\frac{r}{|\zeta|}\right)}}{2\bar{C}_2},
	\label{eq:38}
\end{equation}

Outside the critical point $(r>r_c)$ the minus sign describes the subsonic accretion branch, while inside $(r<r_c)$ the plus sign corresponds to the supersonic branch; the two coincide at $(r=r_c)$ where the square root argument becomes zero. Our study focuses exclusively on the region $r>r_c$.

Figure 9 illustrates the variation of the radial velocity $u^*$ with radial distance r. It can be seen that the radial velocity exhibits a peak at a certain radius outside the black hole event horizon. In Fig.9a, the radial velocity increases with the PFDM parameter $\zeta$. Similarly, Fig.9b shows that the radial velocity also rises as the magnetic charge parameter $a$ increases. Throughout the region outside the event horizon, the radial velocity of the fluid around the charged-PFDM black hole remains consistently and significantly higher than that around a Schwarzschild black hole.

Using Eqs.\eqref{eq:31} and \eqref{eq:38}, we obtain:

{\small
\begin{equation}
	\rho = \left( \frac{2\bar{C}_2^2}{ r^2 \left( r^2 - \sqrt{r^4 - 4\bar{C}_2 \left(1 - \frac{2M}{\sqrt{r^2+a^2}} + \frac{\zeta}{r} \log\frac{r}{|\zeta|}\right)} \right) } \right)^{1+k}.
	\label{eq:39}
\end{equation}}

\begin{figure*}
	a)\includegraphics[width=8.1 cm]{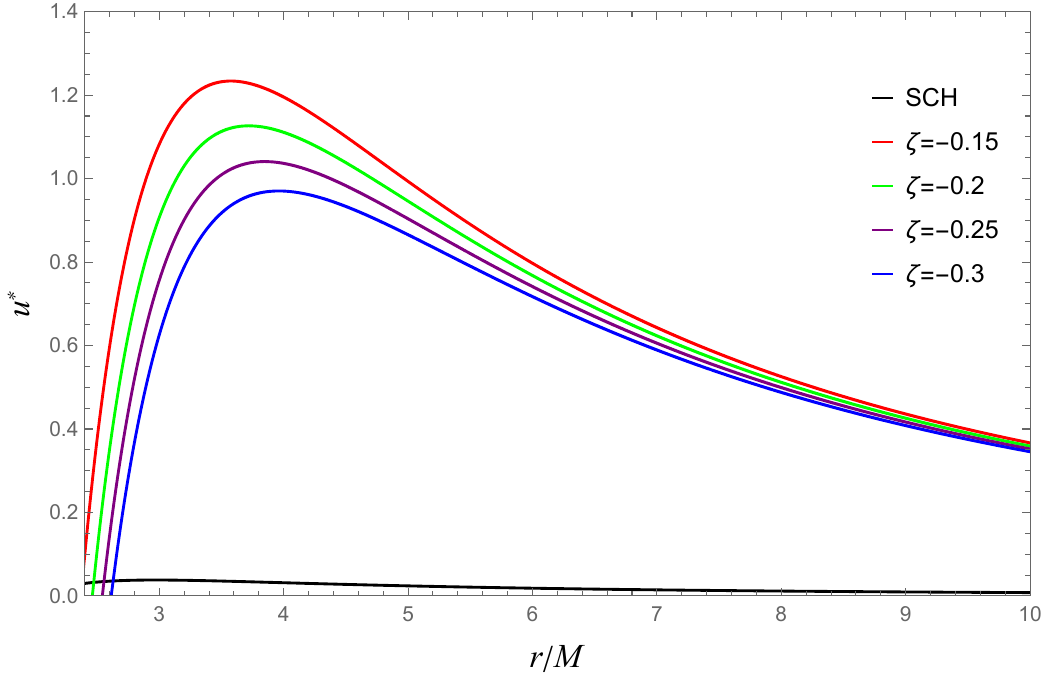}
	b)\includegraphics[width=8.1 cm]{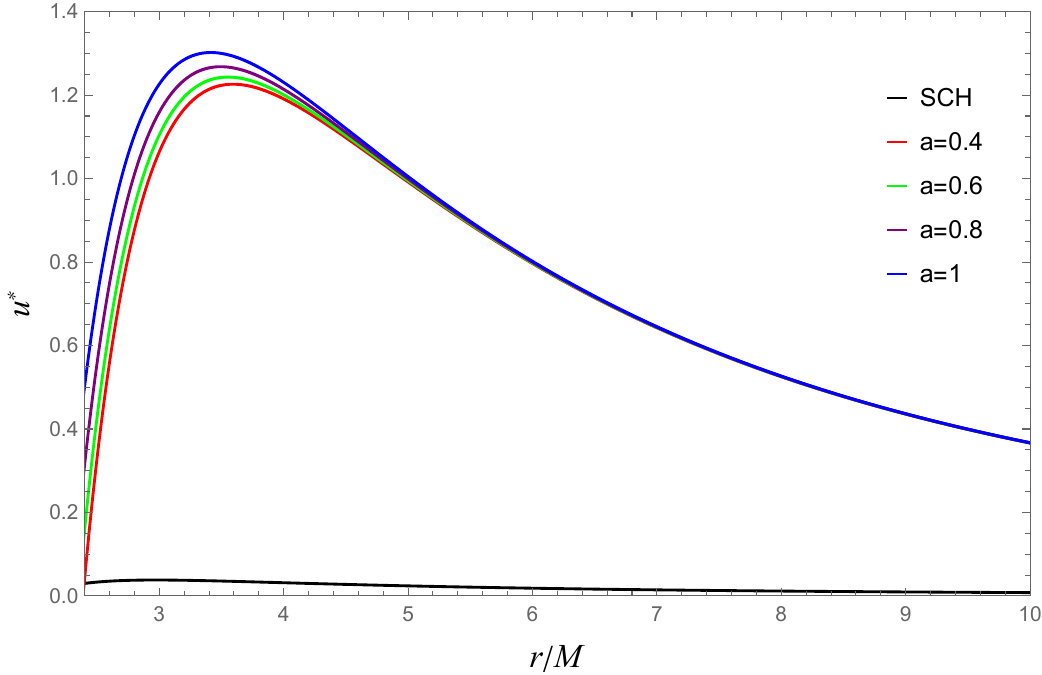}
	\caption{Plots of the fluid density $u^*$ as a function of radius $r$ (a) different values of $\zeta$ with $a=0.5$ (b) different values of $a$ with $\zeta=-0.15$, where the black curve corresponds to the Schwarzschild case.} \label{fig: 6}
\end{figure*}

Figure 10 presents the variation of fluid density $\rho$ with radius $r$ for different values of the PFDM parameter and the magnetic charge parameter. Fig.10a shows that the fluid density decreases with increasing radius, eventually asymptoting to a certain value. The density decreases as the black hole parameters increase, yet remains consistently higher than in the Schwarzschild case. A similar pattern is observed in Fig.10b.

\begin{figure*}
	a)\includegraphics[width=8.1 cm]{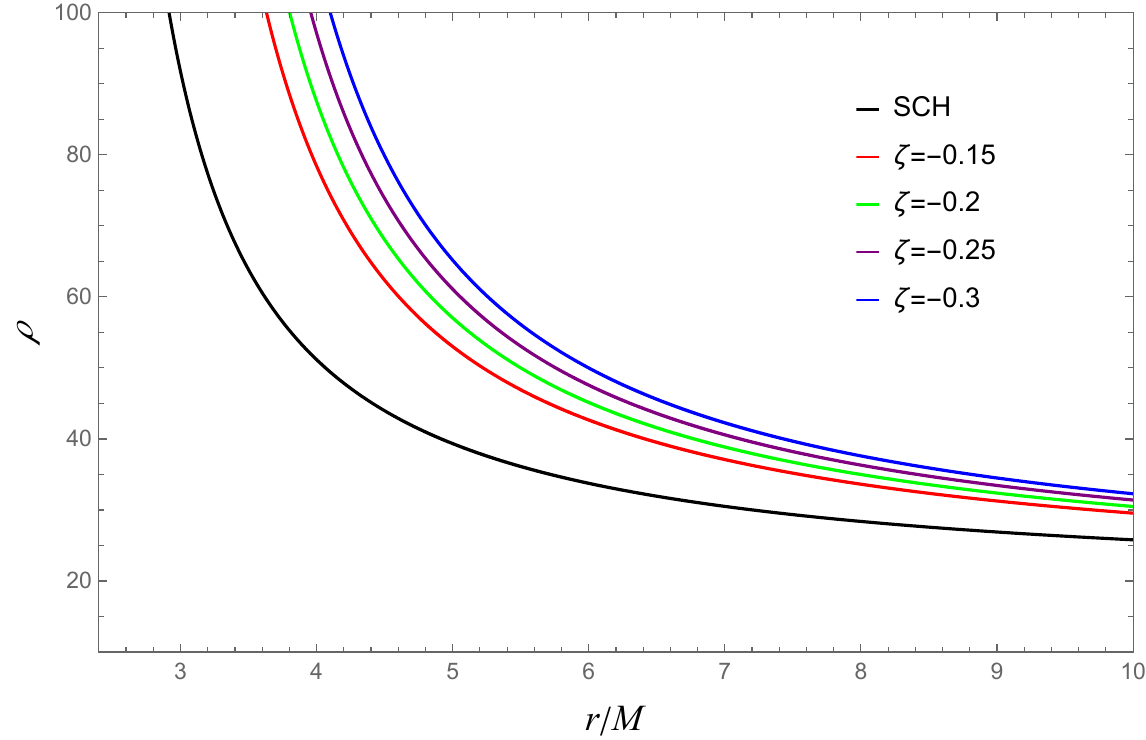}
	b)\includegraphics[width=8.1 cm]{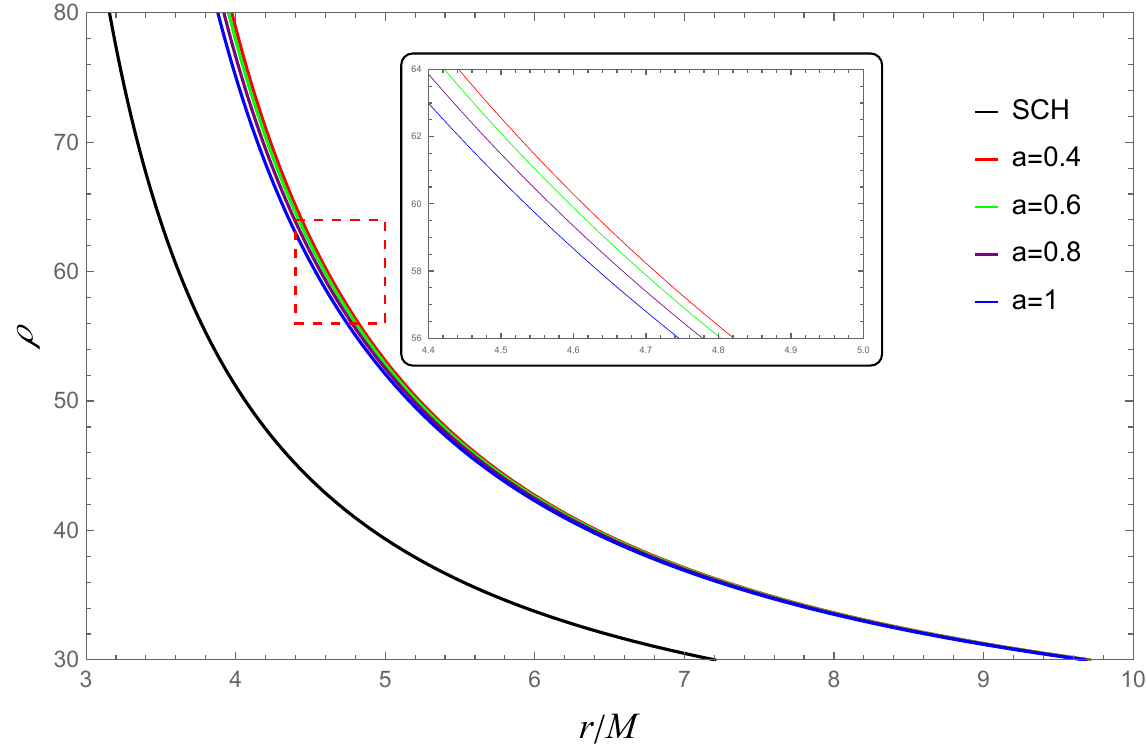}
	\caption{Fluid density $\rho$ versus $r$ with the equation-of-state parameter $k=0.5$ (a) for $a=0.5$, varying $\zeta$, (b) for $\zeta=-0.15$, varying $a$.} \label{fig: 7}
\end{figure*}

Through accretion, the rate of change of the black hole mass is given by $\dot{M} = -4\pi r^2 T_0^r$. Employing Eqs.\eqref{eq:31} and \eqref{eq:32}, the relation takes the form \cite{Babichev:2005py, Shahzad:2024ljt}:

\begin{equation}
	\dot{M} = 4\pi C_2 (\rho_\infty + p_\infty) M^2.
	\label{eq:40}
\end{equation}

In order to clarify the time evolution of the black hole mass, utilizing the initial mass $M_i$ and Eq.\eqref{eq:40}, we present the following:

\begin{equation}
	M_t  = \frac{M_i}{1 - t/t_{\mathrm{cr}}},
	\label{eq:41}
\end{equation}

where $t_{\mathrm{cr}}$ is the characteristic accretion time:

\begin{equation}
	t_{\mathrm{cr}} = [4\pi C_2 (\rho + p) \sqrt{f(r_\infty)} M_i]^{-1}.
	\label{eq:42}
\end{equation}

Combining with Equation \eqref{eq:42}, we find that the BH mass increases to infinity within a finite time when $t=t_{\mathrm{cr}}$. Therefore, the mass accretion rate of the BH is given by \cite{Shahzad:2024ljt}:

\begin{equation}
	\dot{M} = 4\pi \bar{C}_2 (\rho + p) M^2.
	\label{eq:43}
\end{equation}

Figure 11 illustrates the relationship between the mass accretion rate and the radial coordinate $r$. It is observed that the accretion rate decreases outward with increasing black hole radius $r$, and also diminishes as the black hole parameters $a$ or $\zeta$ grow. However, it consistently remains higher than the accretion rate of a Schwarzschild black hole. Furthermore, Fig.11a exhibits a wider spread compared to Fig.11b, indicating a stronger influence of the PFDM parameter.

\begin{figure*}
	a)\includegraphics[width=8.1 cm]{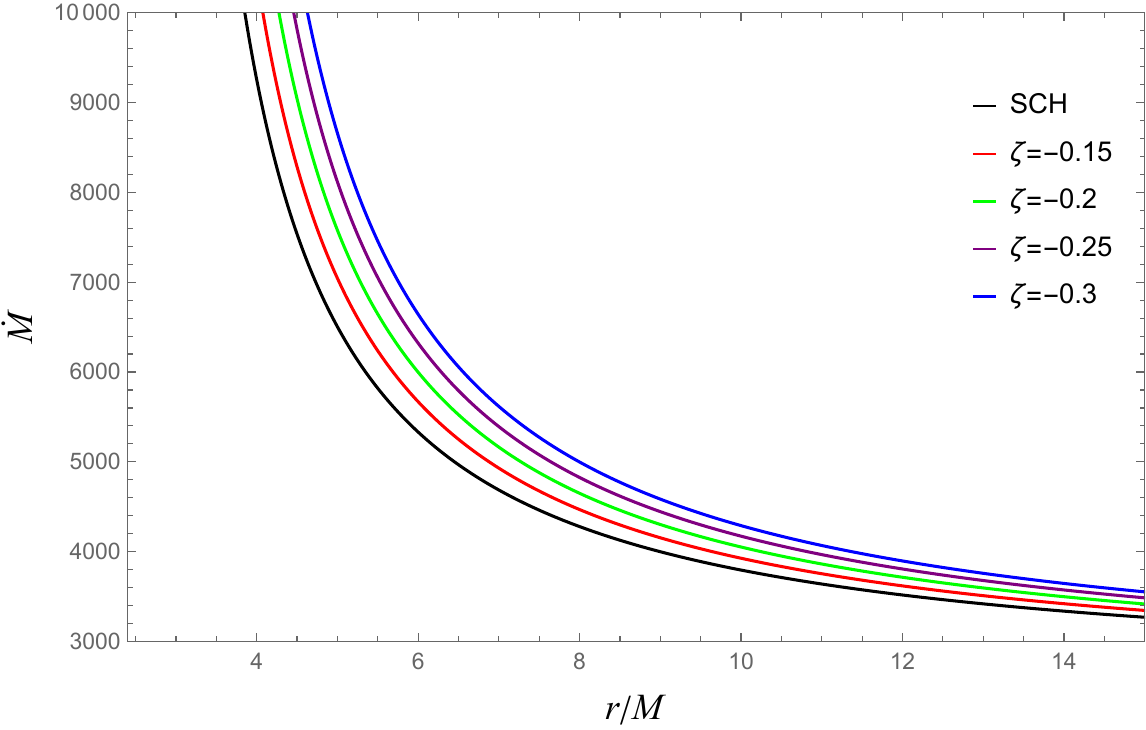}
	b)\includegraphics[width=8.1 cm]{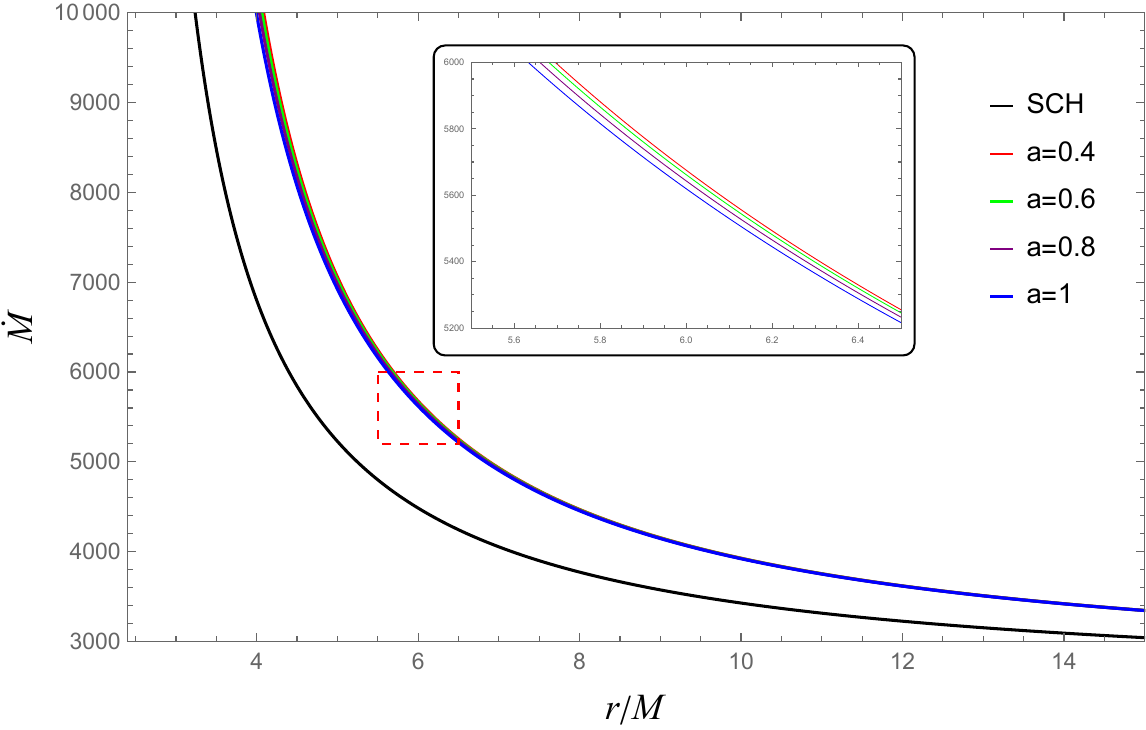}
	\caption{The mass accretion rate $\dot{M}$ as a function of radius $r$ (a) by assuming $a=0.5$ and different values of $\zeta$, (b) for $\zeta=-0.15$ and various values of $a$.} \label{fig: 8}
\end{figure*}

\section{Conclusion}
\label{sec:conclusion}

This work presents a systematic investigation of accretion dynamics around static black holes endowed with a magnetic charge within a perfect fluid dark matter (PFDM) background. We first employed the shadow observations of M87* from the EHT to constrain the black hole’s magnetic charge parameter $a$ and the PFDM parameter $\zeta$. The observationally allowed parameter ranges were determined to be $0 < a \leq 1$ and $-0.3\leq \zeta<0$, thereby establishing the astrophysical viability of this model.

Within the constrained parameter space, we performed a detailed analysis of the orbital dynamics of test particles around the black hole. Our results show that both the magnetic charge and the dark matter environment significantly alter the effective potential, the location of the innermost stable circular orbit (ISCO), and the specific energy and specific angular momentum of the particles. Specifically, a more negative PFDM parameter $\zeta$ (i.e., a stronger dark matter effect) leads to a larger ISCO radius, whereas an increase in the magnetic charge a tends to shift the ISCO inward. These changes in orbital properties directly influence the inner boundary of the accretion disk and the associated energy release processes.

Based on the Novikov–Thorne thin-disk model, we computed the radiative energy flux, temperature profile, and observed spectrum of the accretion disk. We found the local radiative flux $F(r)$ and temperature $T(r)$ at a given radius are lower for the charged-PFDM black hole than for a Schwarzschild black hole, but its overall radiative efficiency $\eta^*$ and total observed luminosity are higher. This is primarily due to two mechanisms: first, a larger ISCO radius (especially for more negative $\zeta$) provides a greater effective radiating area; second, a stronger gravitational potential (especially for larger $a$) enhances the fraction of gravitational energy released during the infall of particles. The radiative efficiency $\eta^*$ can reach up to approximately $7.5\%$, notably higher than the $6\%$ of a Schwarzschild black hole, demonstrating that the magnetic charge and PFDM coupling can significantly enhance the radiative output of thin accretion disks.

In addition to the high-angular-momentum thin disk scenario, we examined the dynamics of spherically symmetric, steady-state accretion, which is more representative of diffuse, low-angular-momentum environments such as dark matter accretion. Our study reveals that in a PFDM background, the fluid’s radial velocity, density profile, and the black hole’s mass accretion rate are all significantly modulated by the parameters $a$ and $\zeta$. This provides a theoretical perspective on how black holes may grow by accreting surrounding material in quiescent galactic environments, complementing the high-luminosity picture provided by thin disk analysis.

By combining these two complementary accretion regimes, this research elucidates the profound influence of the coupling between magnetic charge and PFDM on black hole accretion processes. These effects are imprinted not only on macroscopic scales such as the black hole shadow, but also in the detailed dynamics of the accretion flow and its radiative output. Our results suggest that future joint analyses of black hole shadows, continuum spectral energy distributions, and luminosity measurements hold promise for imposing tighter observational constraints on the microstructure (magnetic charge) of supermassive black holes at galactic centers and their ambient dark matter environments. This could pave a potential path toward exploring new physics beyond General Relativity and standard astrophysical models.

\section*{Acknowledgments}
This research partly supported by the
National Natural Science Foundation of China (Grant No. 12265007).

\newpage


\bibliographystyle{unsrt}  
\bibliography{ref1}
\bibliographystyle{apsrev4-1}

\end{document}